# VIS-MAE: An Efficient Self-supervised Learning Approach on Medical Image Segmentation and Classification

Zelong Liu[1], Andrew Tieu[1], Nikhil Patel[1], George Soultanidis[1], Louisa Deyer[1], Ying Wang[2], Sean Huver[3], Alexander Zhou[1], Yunhao Mei[4], Zahi A. Fayad[1], Timothy Deyer[5,6], and Xueyan Mei[1,7]

[1] BioMedical Engineering and Imaging Institute, Icahn School of Medicine at Mount Sinai, New York, NY, USA
[2] Department of Mathematics, University of Oklahoma, Norman, OK, USA
[3] NVIDIA, Santa Clara, CA, USA
[4] Erasmus University Rotterdam, Rotterdam, The Netherlands
[5] East River Medical Imaging, New York, NY, USA
[6] Department of Radiology, Cornell Medicine, New York, NY, USA
[7] Windreich Department of Artificial Intelligence and Human Health, Icahn School of Medicine at Mount Sinai, New York, NY, USA

**Abstract.** Artificial Intelligence (AI) has the potential to revolutionize diagnosis and segmentation in medical imaging. However, development and clinical implementation face multiple challenges including limited data availability, lack of generalizability, and the necessity to incorporate multi-modal data effectively. A foundation model, which is a large-scale pre-trained AI model, offers a versatile base that can be adapted to a variety of specific tasks and contexts. Here, we present **VI**sualization and **S**egmentation **M**asked **A**uto**E**ncoder (VIS-MAE), novel model weights specifically designed for medical imaging. Specifically, VIS-MAE is trained on a dataset of 2.5 million unlabeled images from various modalities (CT, MR, PET, X-rays, and ultrasound), using self-supervised learning techniques. It is then adapted to classification and segmentation tasks using explicit labels. VIS-MAE has high label efficiency, outperforming several benchmark models in both in-domain and out-of-domain applications. In addition, VIS-MAE has improved label efficiency as it can achieve similar performance to other models with a reduced amount of labeled training data (50% or 80%) compared to other pre-trained weights. VIS-MAE represents a significant advancement in medical imaging AI, offering a generalizable and robust solution for improving segmentation and classification tasks while reducing the data annotation workload. The source code of this work is available at https://github.com/lzl199704/VIS-MAE.

## 1 Introduction

Recent advances in the field of artificial intelligence (AI) have given rise to the development of foundation models, machine learning models that are trained on large, diverse datasets that can be adapted to a wide variety of downstream tasks [1]. While traditional deep learning models are specifically trained for designated applications, such as the classification of interstitial lung disease [2] or COVID-19 [3], and consistently underperform when repurposed for other tasks [4], foundation models offer more generalized and adaptable capabilities. Following initial training, foundation models can be fine-tuned to numerous different tasks. In the field of medical imaging, foundation models have a wide range of potential applications such as improved diagnostic accuracy, increased efficiency of reads, and prediction of disease outcomes. The incorporation of foundation models into clinical workflows has the potential to revolutionize the field of healthcare delivery. Despite these advances, there remain several challenges in the development of foundation models. Training of foundation models through traditional supervised learning approaches is limited by the need for large quantities of labeled data, which is often a prohibitively time-consuming and cost-intensive process [5]. This is further compounded by an overall lack of high-quality and open-source medical imaging data [6–8], hindering the generalizability and accuracy of developed models [9, 10].

The Segment Anything Model (SAM) [11], a foundation model trained on a dataset of 11 million natural images with over 1 billion image masks, showcased the ability to automatically segment any image using prompts. Soon after, this capability was applied to medical images with the release of MedSAM [12], trained on a large-scale medical imaging dataset with over 1 million medical image-mask pairs across a wide array of modalities and protocols, thus opening the door to many possibilities in the field of medical imaging analysis. Following MedSAM's release, increasing numbers of medical imaging foundation models have been reported for a variety of applications. RETFound [13], a foundation model trained on 1.3 million retinal images, offers generalizable capabilities to detect multiple retinal diseases. Other medical foundation models can perform segmentation of moving structures [14], volumetric segmentation [15], diagnosis and prognosis of ocular disease [16], and assessment of clinical pathology images [17].

In response to the lack of high quality open-source medical imaging data, current approaches to medical imaging analysis have adopted the use of self-supervised learning (SSL), a training method in which models are able to infer labels through latent features of unlabeled data. The success of SSL relies on pretext tasks designed on unlabeled data, which can learn image representations through reconstruction, jigsaw puzzle solving, or contrastive learning [18]. SSL methods such as a simple framework for contrastive learning of visual representations (SimCLR) [19] and Masked Autoencoder [20] have been shown to achieve comparable results or even outperform models using supervised learning methods. The architecture of SimCLR can employ a simple and efficient backbone and a classifier head, and SimCLR has been shown to be an effective pre-training strategy for medical image classification [21]. However, since SimCLR only provides pre-trained encoder weights, while an MAE could offer pre-trained weights for both the encoder and decoder, an MAE could be more beneficial for both segmentation and classification. Recently, Swin MAE [22], a masked autoencoder using Swin Transformer as its backbone, demonstrated the feasibility of achieving such results on smaller datasets without the use of pre-trained models.

To improve the training efficiency and model generalizability, we introduce a self-supervised learning model with weights trained on a large-scale database, the **VI**sualization and **S**egmentation **M**asked **A**uto**E**ncoder (VIS-MAE) model. This model employs a Swin Transformer-based masked autoencoder, designed for a wide range of downstream tasks. The pre-trained weights of VIS-MAE were developed using a dataset of 2.5 million images from various imaging modalities, including CT, MRI, PET/CT, radiography (X-ray), and ultrasound (US). The generic VIS-MAE (VIS-MAE-Generic) weights were trained on the entire dataset to ensure broad applicability across different medical imaging types, while the modality-specific VIS-MAE (VIS-MAE-Modality) weights were trained from images from individual modalities, enhancing their ability to identify modality-specific anatomical and pathological features. Our study shows that VIS-MAE can outperform traditional supervised models in terms of performance and generalizability across diverse medical imaging tasks, while also reducing the dependency on extensively labeled datasets.

## 2 Methodology

### 2.1 VIS-MAE Model Development

As shown in Fig. 1, our VIS-MAE was created by incorporating Swin MAE [22] with an additional segmentation decoder or classification layer. VIS-MAE incorporates a modified version of Swin Transformer [23] as its backbone, while a window masking method was applied during the MAE process. The modification includes the same block number in each stage to reduce the trainable parameters and improve training efficiency. Each Swin Transformer block contains a LayerNorm (LN) layer, a window-based multi-head self-attention (W-MSA) module, a residual connection and a multilayer perceptron (MLP) layer with GELU non-linearity activation. The output of $l^{th}$ Swin Transformer block layer, $z_l$, can be annotated as:

$$\hat{z}_l = \text{W} - \text{MSA}\left(\text{LN}\left(z_{l-1}\right)\right) + z_{l-1} \tag{1}$$

$$z_l = \text{MLP}\left(\text{LN}\left(\hat{z}_l\right)\right) + \hat{z}_l \tag{2}$$

Equations 1 and 2 illustrate the feature extraction process in VIS-MAE. Each attention stage is wrapped with a residual connection, following the Swin Transformer layers to facilitate deeper network architectures without degradation in performance. The output from the W-MSA is summed with its input to preserve the flow of gradients and prevent the vanishing gradient problem.

We also modified the masking method in VIS-MAE because Swin Transformer blocks use a 4 × 4 patch size which leads to easy reconstruction for deep learning models. In our study, the minimum size (16 × 16 pixels) of a random mask will contain multiple image patches. Masking could encompass up to 75% of the original image. The objective function of the VIS-MAE encourages the model to minimize the reconstruction error between the original image patches $\boldsymbol{X}$ and the reconstructed image patches $\bar{\boldsymbol{X}}$, which are generated from the masked version of the original image through the encoder-decoder architecture of the VIS-MAE:

$$Loss = \| X - x \|_2^2 \tag{3}$$

In Eq. 3, the squared L2 norm function will quantify the pixel-wise reconstruction error across the masked regions of ground truth and the predicted images. The integration of Swin Transformer with the modified masking strategy allows VIS-MAE to effectively handle diverse and complex large-scale training datasets, providing a powerful tool for developing pre-trained model weights for precise image segmentation and classification. In addition, the encoder-decoder architecture of VIS-MAE can be easily tailored to the needs of visual understanding tasks by adding a skip connection layer or classification head, enabling our model to achieve state-of-the-art performance on standard benchmarks.

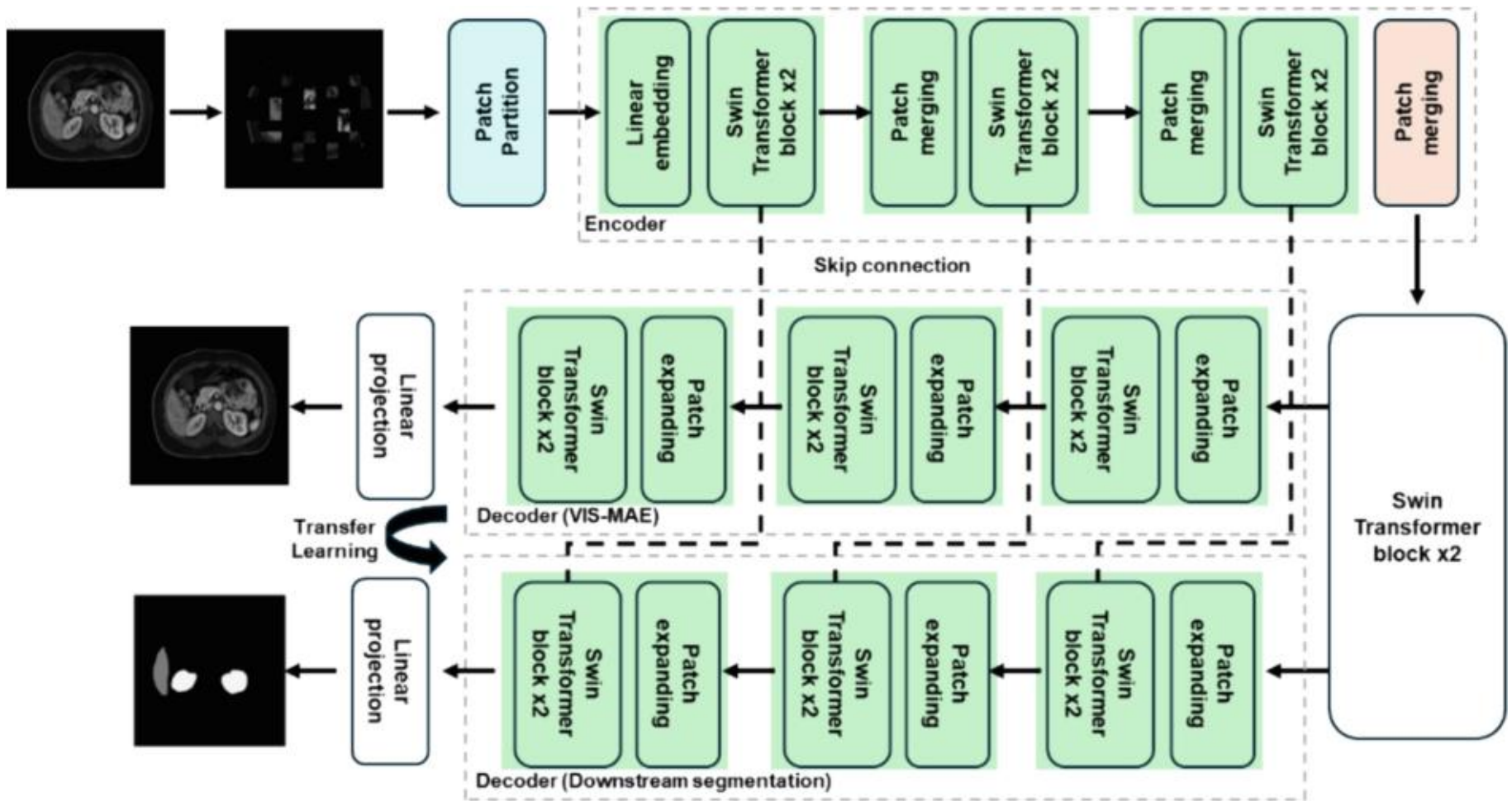

**Fig. 1.** The architecture of VIS-MAE and its implication on downstream tasks, including segmentation. VIS-MAE consists of an encoder, bottleneck blocks, and a decoder. Images are first masked randomly with a patch, and then the network reconstructs the original image. The VIS-MAE model weights can then be efficiently adapted to downstream applications and have additional skip connection layers added for segmentation tasks.

## 3 Experiments and Results

### 3.1 Datasets

The dataset used to develop VIS-MAE was collected from an outpatient radiology practice (RadImageNet LLC) in New York between 2005 and 2022. It consists of 2,486,425 images from five imaging modalities: MR (1,199,904 images), CT (570,943 images), PET/CT (65,731 images), X-rays (438,521 images), and ultrasound (211,325 images).

This dataset expands the RadImageNet [24] project, adding new modalities and anatomical regions. MR, CT, and PET/CT images were specifically selected for the presence of significant pathology by a radiologist, while all images from X-ray and ultrasound studies were included. The patient cohort had a mean age of 54 (SD = 18), with 191,193 female and 169,915 male participants; 175 did not report gender. Sampled ethnicity data indicate 18.73% Hispanic or Latino and 81.27% not, while race was 70.93% White, 14.78% Black or African American, 7.26% Asian, and small percentages for other categories. All demographic data were self-reported. All images were resized to 224 × 224 pixels and normalized to 0–1 for model development. For CT, before image transformation all images had a window/level applied as described in the corresponding DICOM header.

For downstream tasks, we assessed VIS-MAE pre-trained weights on eight segmentation datasets, including the BTCV-Abdomen dataset [25] with 2,178 CT slices of 13 abdominal organs, the ACDC dataset [26] with 2,978 cardiac MR slices of heart regions, the AMOS dataset [27] with 2,476 abdomen MR slices of 12 organs, a prostate dataset [28] with 3,554 MR slices, a brain segmentation dataset with 1,373 MR slices for glioma [29], the BUSI segmentation dataset [30] of 647 breast ultrasound images, the Thyroid Ultrasound Cine-clip (TUCC) [31] dataset with 17,641 thyroid ultrasound images, and the ISIC [32] dataset with 1,279 dermoscopy images.

Additionally, VIS-MAE was evaluated on six classification tasks: a COVID-19 dataset [3] of 9,050 CT chest images, an internal sarcoidosis dataset of 1,231 PET/MR images collected from the Mount Sinai Hospital, the BUSI dataset [29] of 647 images for breast lesion malignancy detection, an ACL dataset [33] of 1,021 MR knee images, a knee osteoarthritis dataset [34] of 1,650 X-ray images with five grades, and the NIH Chest X-ray [35] dataset with 112,120 images of 14 pulmonary classes.

### 3.2 VIS-MAE Implementation Details

**Upstream Model Development.** The VIS-MAE model training was extensive, developing distinct pre-trained weights for different imaging modalities over 800 epochs with a batch size of 640, using the AdamW optimizer and a mean square error loss function. The initial learning rate was 0.0001, decreasing with training. There was a 10-epoch warmup. Training took between 11 and 516 h on 8 NVIDIA DGX A100 GPUs, depending on the modality. To explore modality-specific features and their comparison to a unified model, we developed two versions of VIS-MAE: VIS-MAE-Modality for specific imaging modalities and VIS-MAE-Generic as a comprehensive model. The VIS-MAE-Generic was created using a dataset comprised of 2.5 million images from five imaging modalities. In contrast, VIS-MAE-Modality consists of five models: VIS-MAE-MR, VIS-MAE-PET, VIS-MAE-CT, VIS-MAE-XRAY, and VIS-MAE-ultrasound, designed to address the nuances of specific imaging modalities.

**Downstream Model Development.** To evaluate the utility of VIS-MAE for various medical applications, eight segmentation and six classification downstream tasks were chosen based on their anatomical, modal, and pathological diversity. The VIS-MAE pre-trained weights were finetuned on these downstream applications with the following strategies. VIS-MAE models had standardized warmup periods (40 epochs for segmentation, 10 for classification) and training durations varied (150 epochs for segmentation, 50 for classification), with adjustable learning rates and batch sizes. We first evaluated the performance of the modality-specific VIS-MAE models against the VIS-MAE-Generic model. In addition, we compared VIS-MAE against several benchmark models in both segmentation and classification tasks. For segmentation, we utilized strategies including nnU-Net [36], TransUNet [37], and SimCLR. nnU-Net was adapted to match the VIS-MAE data distribution and underwent a similar training regime, utilizing a unique pre-processing method tailored to enhance performance. TransUNet employed the SGD optimizer, leveraging pre-trained weights from ImageNet-1k [38], a vast dataset of approximately 1.3 million natural images across 1,000 categories, widely used for training models via traditional supervised learning. Both nnU-Net and TransUNet were configured using the default parameters recommended in their original publications. We also used models pre-trained on RadImageNet, a specialized radiological dataset featuring 1.35 million images annotated across 165 pathological labels and 14 anatomical regions, including CT, MR, and ultrasound modalities. Both RadImageNet and ImageNet-1k pre-trained weights were configured to follow similar downstream parameters as VIS-MAE for consistency, allowing for a direct comparison of performance. Additionally, the SimCLR weights were pre-trained using a Swin Transformer architecture with the same data as VIS-MAE, ensuring aligned training parameters for subsequent tasks. Except for nnU-Net and TransUNet, all models employed distinct pre-trained weights for segmentation and classification but utilized the same architectural framework and fine-tuning processes for downstream applications. Models were evaluated with the Dice Similarity Coefficient (DSC), precision, and recall for segmentation tasks, and with the area under the ROC curve (ROC AUC) and Precision-Recall curve AUC (PR AUC) score for classification tasks. For each dataset containing fewer than 4000 images, we conducted five-fold cross-validation and reported the average metrics from these five folds. In developing the segmentation models, we employed a combination of the dice loss and cross-entropy loss in a 4:6 ratio. For training classification models, we utilized cross-entropy loss.

### 3.3 VIS-MAE on Segmentation and Classification Datasets

The VIS-MAEs models were assessed across eight medical image segmentation datasets with various imaging modalities and anatomies. They were compared to several benchmarks, including nnU-Net, TransUNet, two supervised learning pre-trained Swin Transformer weights, RadImageNet and ImageNet, and another SSL strategy, SimCLR. VIS-MAE showed superior or comparable performance to these models, as illustrated in Tables 1a and 1b. Additionally, VIS-MAE's performance on classification datasets were compared to RadImageNet, ImageNet, and SimCLR weights and outperformed or achieved equivalent performance to these existing models, with detailed outcomes presented in Table 2.

**Table 1a.** Model performance on segmentation datasets in DSC, precision, and recall.

| Methods | BTCV (CT) | | | ACDC (MR) | | | AMOS (MR) | | | Glioma (MR) | | |
|---|---|---|---|---|---|---|---|---|---|---|---|---|
| | DSC | Precision | Recall | DSC | Precision | Recall | DSC | Precision | Recall | DSC | Precision | Recall |
| VIS-MAE-Modality | 0.844 | 0.809 | 0.765 | 0.879 | 0.893 | 0.878 | 0.891 | 0.899 | 0.875 | 0.866 | 0.881 | 0.875 |
| VIS-MAE-Generic | 0.854 | 0.815 | 0.781 | 0.882 | 0.895 | 0.877 | 0.862 | 0.903 | 0.869 | 0.871 | 0.870 | 0.887 |
| nnU-Net | 0.842 | 0.803 | 0.761 | 0.883 | 0.899 | 0.873 | 0.877 | 0.847 | 0.823 | 0.881 | 0.889 | 0.890 |
| TransUNet | 0.858 | 0.794 | 0.748 | 0.874 | 0.895 | 0.855 | 0.863 | 0.897 | 0.836 | 0.833 | 0.856 | 0.840 |
| RadImageNet | 0.852 | 0.831 | 0.790 | 0.880 | 0.891 | 0.875 | 0.873 | 0.899 | 0.875 | 0.874 | 0.878 | 0.881 |
| ImageNet | 0.855 | 0.820 | 0.781 | 0.882 | 0.895 | 0.879 | 0.869 | 0.893 | 0.892 | 0.874 | 0.881 | 0.886 |
| SimCLR | 0.813 | 0.768 | 0.726 | 0.861 | 0.881 | 0.853 | 0.847 | 0.860 | 0.851 | 0.852 | 0.864 | 0.862 |

**Table 1b.** Model performance on segmentation datasets in DSC, precision, and recall.

| Methods | Prostate (MR) | | | TUCC (US) | | | BUSI (US) | | | ISIC (Dermoscopy) | | |
|---|---|---|---|---|---|---|---|---|---|---|---|---|
| | DSC | Precision | Recall | DSC | Precision | Recall | DSC | Precision | Recall | DSC | Precision | Recall |
| VIS-MAE-Modality | 0.902 | 0.840 | 0.824 | 0.720 | 0.740 | 0.717 | 0.780 | 0.797 | 0.791 | 0.923 | 0.926 | 0.906 |
| VIS-MAE-Generic | 0.825 | 0.834 | 0.816 | 0.706 | 0.687 | 0.786 | 0.775 | 0.789 | 0.794 | 0.908 | 0.920 | 0.915 |
| nnU-Net | 0.833 | 0.837 | 0.833 | 0.716 | 0.765 | 0.756 | 0.783 | 0.803 | 0.811 | 0.904 | 0.900 | 0.930 |
| TransUNet | 0.815 | 0.839 | 0.797 | 0.696 | 0.788 | 0.704 | 0.776 | 0.769 | 0.747 | 0.919 | 0.936 | 0.918 |
| RadImageNet | 0.829 | 0.830 | 0.833 | 0.705 | 0.759 | 0.758 | 0.780 | 0.805 | 0.814 | 0.916 | 0.928 | 0.912 |
| ImageNet | 0.832 | 0.844 | 0.832 | 0.718 | 0.736 | 0.799 | 0.788 | 0.810 | 0.822 | 0.919 | 0.930 | 0.921 |
| SimCLR | 0.802 | 0.816 | 0.803 | 0.619 | 0.708 | 0.723 | 0.758 | 0.781 | 0.781 | 0.910 | 0.923 | 0.914 |

**Table 2.** VIS-MAE performance on classification datasets in ROC AUC score and PR AUC score.

| Datasets | COVID-19 (CT) | | Sarcoidosis (PET/MR) | | ACL tear (MR) | | Knee osteoarthritis (X-ray) | | BUSI (US) | | NIH Chest (X-Ray) | |
|---|---|---|---|---|---|---|---|---|---|---|---|---|
| | ROC AUC | PR AUC | ROC AUC | PR AUC | ROC AUC | PR AUC | ROC AUC | PR AUC | ROC AUC | PR AUC | ROC AUC | PR AUC |
| VIS-MAE-Modality | 0.857 | 0.856 | 0.624 | 0.237 | 0.946 | 0.958 | 0.938 | 0.797 | 0.867 | 0.794 | 0.802 | 0.262 |
| VIS-MAE-Generic | 0.862 | 0.867 | 0.680 | 0.338 | 0.936 | 0.953 | 0.800 | 0.531 | 0.736 | 0.591 | 0.799 | 0.256 |
| RaImageNet | 0.867 | 0.876 | 0.637 | 0.284 | 0.938 | 0.950 | 0.911 | 0.735 | 0.868 | 0.806 | 0.804 | 0.269 |
| ImageNet | 0.863 | 0.861 | 0.616 | 0.234 | 0.937 | 0.956 | 0.926 | 0.766 | 0.787 | 0.658 | 0.812 | 0.275 |
| SimCLR | 0.819 | 0.827 | 0.644 | 0.231 | 0.912 | 0.931 | 0.898 | 0.707 | 0.899 | 0.841 | 0.749 | 0.189 |

### 3.4 Label Efficiency in Segmentation and Classification Applications

Label efficiency measures the training data and annotations required to achieve desired performance, reflecting the burden on medical professionals. To assess the label efficiency of VIS-MAE, we compared its performance in segmentation and classification datasets using only 5%, 10%, 25%, 50%, and 80% of the original training data. VIS-MAE's performance across various datasets compared to other models was demonstrated in Figs. 2 and 3.

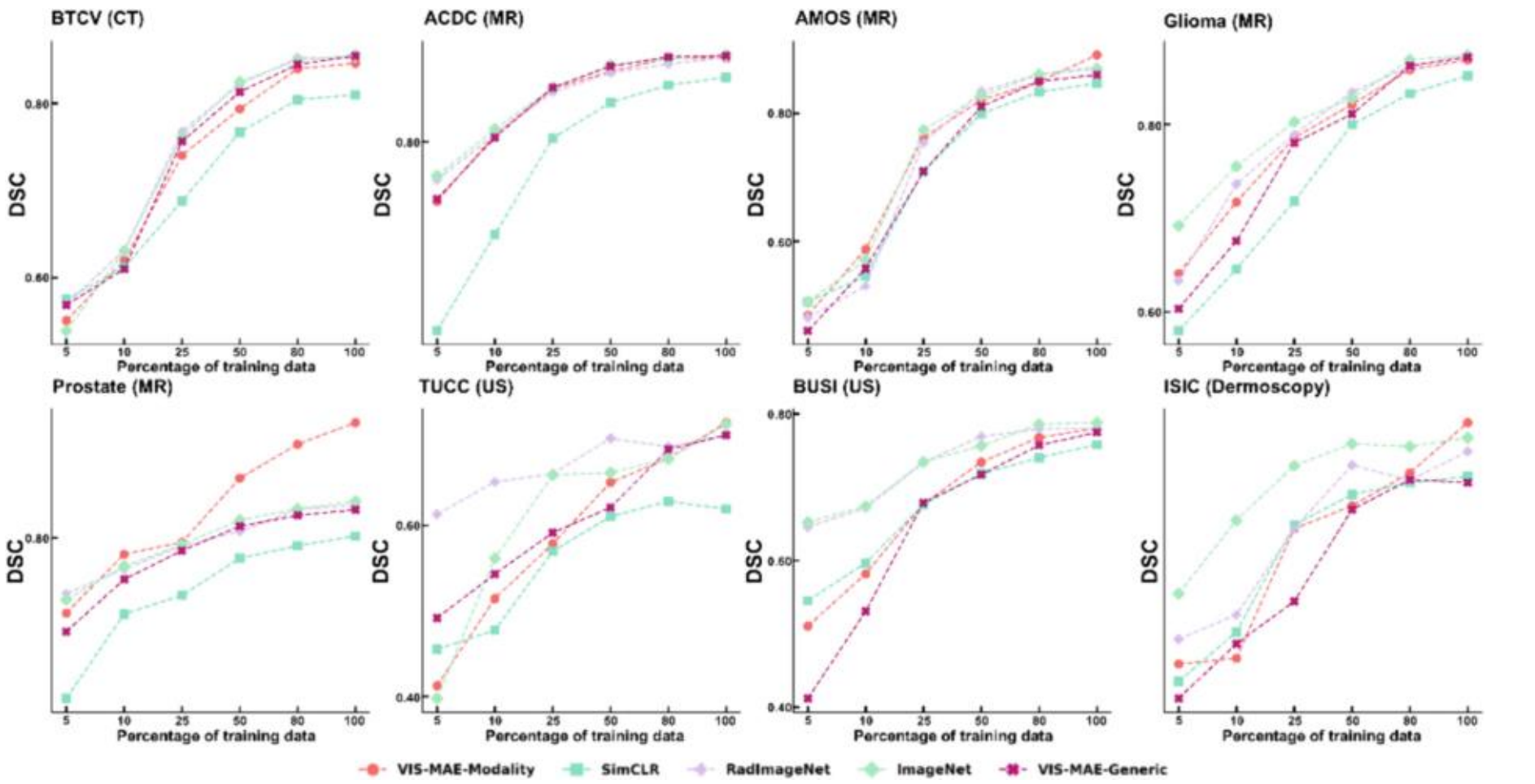

**Fig. 2.** Label efficiency of pre-trained weights VIS-MAE-Modality, VIS-MAE-Generic, SimCLR, RadImageNet and ImageNet on eight segmentation datasets, using different percentages of training data. All models were trained using 5%, 10%, 25%, 50%, 80%, and 100% of the training data in each segmentation dataset, and the DSCs were reported.

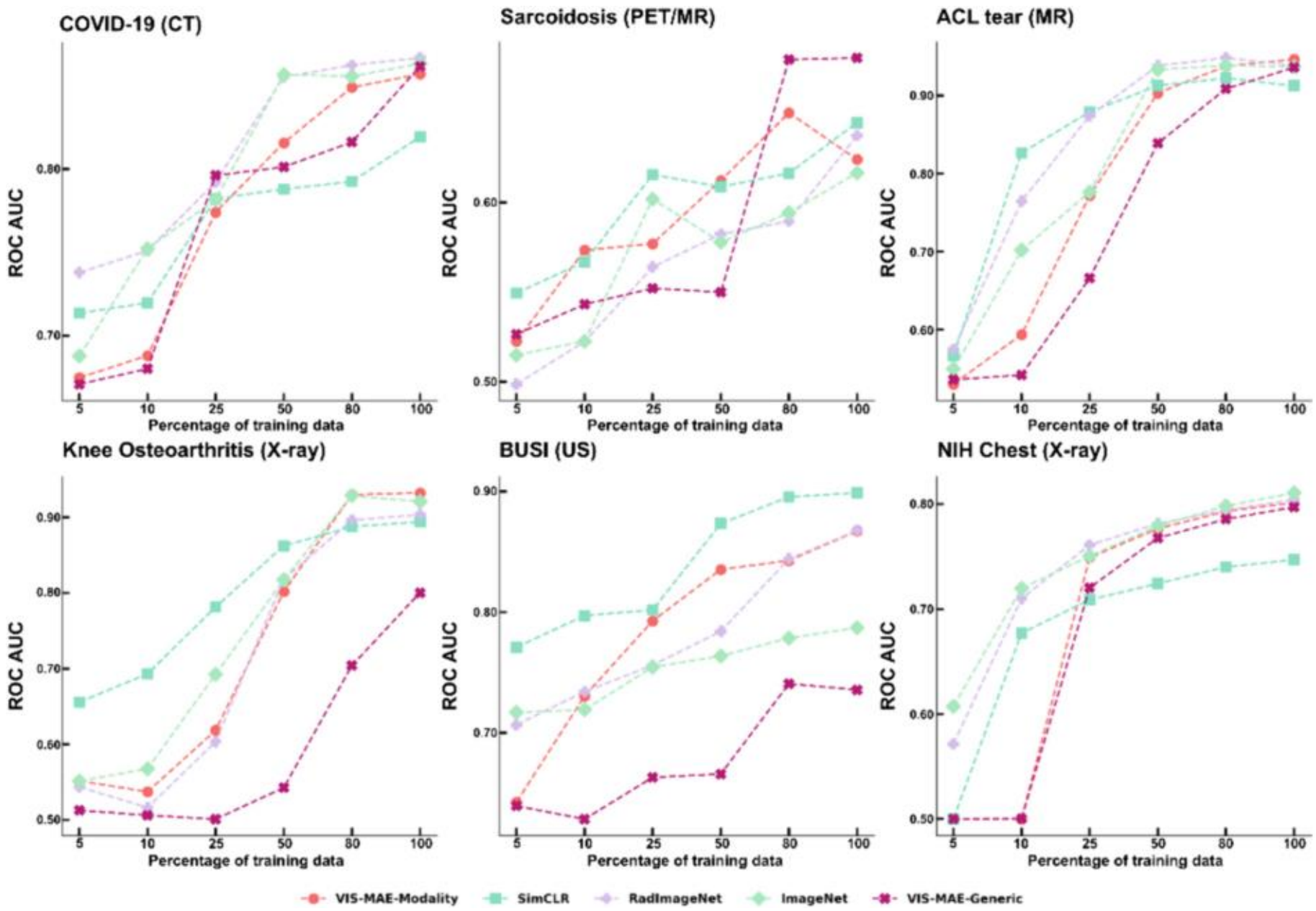

**Fig. 3.** Label efficiency of pre-trained weights VIS-MAE-Modality, VIS-MAE-Generic, SimCLR, RadImageNet and ImageNet on six classification datasets, using different percentages of training data. All models were trained using 5%, 10%, 25%, 50%, 80%, and 100% of the training data in each classification dataset, and the ROC AUC scores were reported.

## 4 Conclusion

This paper presents VIS-MAE, a self-supervised learning model utilizing a masked autoencoder trained on a large collection of radiological images. VIS-MAE's model weights enhance various medical imaging tasks by either integrating skip connections for segmentation or substituting the decoder with a classification layer. Our extensive evaluation demonstrates VIS-MAE's superior performance across multiple imaging modalities, improving both segmentation and classification. Moreover, VIS-MAE enhances label efficiency, reducing the need for extensive annotated data while maintaining high performance and generalizability in medical imaging.

**Acknowledgement.** X.M. is supported by the Eric and Wendy Schmidt AI in Human Health Program.

**Disclosure of Interests.** T.D. is the managing partner of RadImageNet LLC. X.M. has been a paid consultant to RadImageNet LLC.